\documentclass[aps, prd, reprint, superscriptaddress, nofootinbib, bibnotes]{revtex4-2}

\usepackage{graphicx, subfigure}        
\usepackage{aas_macros} 
\usepackage{amssymb,amsmath,placeins,epsfig,array,bm}
\usepackage[usenames,dvipsnames]{color}
\usepackage[normalem]{ulem}
\usepackage[breaklinks,colorlinks,urlcolor=blue,citecolor=blue,linkcolor=blue]{hyperref}
\usepackage{float}
\usepackage[T1]{fontenc}
\usepackage{times}
\usepackage{soul}
\usepackage[capitalise]{cleveref}
\usepackage{bm}
\usepackage{xcolor}
\usepackage{color,units}
\usepackage{aas_macros}

\newcommand{\prob}{\ensuremath{{\rm P}}}

\newcommand{\mbh}{m_{\rm BH}}
\newcommand{\mns}{m_{\rm NS}}
\newcommand{\mchirp}{{\cal{M}}}

\newcommand{\uniform}{{\rm U}}
\newcommand{\tobs}{t_{\rm obs}}
\newcommand{\fobs}{\Delta_{\rm obs}}
\newcommand{\rate}{\Gamma}

\newcommand{\yrgpc}{\ensuremath{{\rm yr^{-1}\,Gpc^{-3}}}}

\newcommand{\dlmax}{\ensuremath{D_{L, \rm{max}}}}
\newcommand{\radconstraint}{\ensuremath{{13.18}\pm 0.11}}
\newcommand{\lambdaconstraint}{\ensuremath{{697.58}_{-27.17}^{+41.17}}}
\newcommand{\waveform}{\texttt{SEOBNRv4\_ROM\_NRTidalv2\_NSBH}}
\newcommand{\program}[1]{\textsc{#1}}

\newcommand{\citeg}[1]{\citep[e.g.,][]{#1}}
\newcommand{\bilby}{{\sc Bilby}}

\begin{document}

\title{Measuring the nuclear equation of state with neutron star-black hole mergers}

\author{Nikhil Sarin}
\affiliation{Nordita,
Stockholm University and KTH Royal Institute of Technology
Hannes Alfvéns väg 12, SE-106 91 Stockholm, Sweden}
\affiliation{Oskar Klein Centre for Cosmoparticle Physics, Department of Physics,
Stockholm University, AlbaNova, Stockholm SE-106 91, Sweden}
\author{Hiranya V. Peiris}
\affiliation{Oskar Klein Centre for Cosmoparticle Physics, Department of Physics,
Stockholm University, AlbaNova, Stockholm SE-106 91, Sweden}
\affiliation{Institute of Astronomy, University of Cambridge, Madingley Road, Cambridge CB3 0HA, UK}
\affiliation{Kavli Institute for Cosmology Cambridge, Madingley Road, Cambridge CB3 0HA, UK}
\author{Daniel J. Mortlock}
\affiliation{Astrophysics Group, Imperial College London, Blackett Laboratory, Prince Consort Road, London SW7 2AZ, UK}
\affiliation{Department of Mathematics, Imperial College London, London SW7 2AZ, UK}
\affiliation{Department of Physics, Stockholm University, AlbaNova, SE-10691 Stockholm, Sweden}
\author{Justin Alsing}
\affiliation{Department of Physics, Stockholm University, AlbaNova, SE-10691 Stockholm, Sweden}
\author{Samaya M. Nissanke}
\affiliation{GRAPPA, Anton Pannekoek Institute for Astronomy and Institute of High-Energy Physics, University of Amsterdam, Science Park 904, 1098 XH Amsterdam, The Netherlands}
\affiliation{Nikhef, Science Park 105, 1098 XG Amsterdam, The Netherlands}
\author{Stephen M. Feeney}
\affiliation{Department of Physics \& Astronomy, University College London, Gower Street, London WC1E 6BT, UK}

\newcommand{\samaya}[1]{{\color{blue}{Samaya: \textbf\small{#1}}}}

\date{\today}
\begin{abstract}
\noindent 
Gravitational-wave (GW) observations of neutron star-black hole (NSBH) mergers are sensitive to the nuclear equation of state (EOS). We present a new methodology for EOS inference with non-parametric Gaussian process (GP) priors, enabling direct constraints on the pressure at specific densities and the length-scale of correlations on the EOS. Using realistic simulations of NSBH mergers, incorporating both GW and electromagnetic (EM) selection to ensure sample purity, we find that a GW detector network operating at O5-sensitivities will constrain the radius of a $\unit[1.4]{M_{\odot}}$ NS and the maximum NS mass with $1.6\%$ and $13\%$ precision, respectively. With the same sample, the projected constraint on the length-scale of correlations in the EOS is $\geq~\unit[3.2]{MeV~fm^{-3}}$. These results demonstrate strong potential for insights into the nuclear EOS from NSBH systems, provided they are robustly identified. 
\end{abstract}

\maketitle

\section{Introduction}
A key aim of modern physics is to understand the behavior of nuclear matter at high densities, and in particular the nuclear equation of state (EOS). However, constraints above the nuclear saturation density are currently beyond the realm of terrestrial experiments~\citeg{Baym2018, Sorensen2023}. For the moment, such progress can only come from observations of extreme astrophysical systems, such as neutron stars (NSs)~\citeg{Lattimer2001, Greif2020}.

There are several distinct astronomical probes of NS physics, including electromagnetic (EM) observations in the radio~\citeg{Demorest2010, Cromartie2020} and X-rays, such as those being made by the NS Interior Composition Explorer (NICER) mission~\citeg{Riley2019}, as well as gravitational wave (GW) observations. The latter possibility was first demonstrated by the multi-messenger GW and EM observations of the binary neutron star (BNS) merger GW170817, which directly measured the NS tidal deformability~\citep{abbott17_gw170817_detection, abbott17_GW170817_mma, abbott19_GW170817_properties, Abbott2018_mr}. These constraints will improve as more BNS mergers are identified and characterized, with projected constraints on the radius of a $\unit[1.4]{M_\odot}$ NS on the order of a few percent with gravitational-wave detectors operating at design sensitivity~\citep{HernandezVivanco2019, Wysocki2020, Landry2020, Ghosh2022, Finstad2022}. However, the expected rate of new discoveries is highly uncertain~\citep{gwtc3pop}. 

The GW emissions produced by neutron star-black hole (NSBH) mergers are also sensitive to the NS tidal deformability~\citep{Huang2021, Clarke2023}, providing a distinct way of measuring both the high-density nuclear EOS and the BH and NS mass and spin distributions~\citep{Abbott2021_nsbh}. Importantly, NSBH systems have higher total masses than BNSs and so produce stronger GW signals that are detectable at considerably greater distances. And, while the rate of NSBH mergers is also highly uncertain~\citep{gwtc3pop}, it is possible that they could come to dominate over BNS mergers in terms of detected numbers. If so, NSBH mergers could potentially provide the best constraints on the high-density nuclear EOS, a possibility we explore here.

We begin by describing our simulations, including the combined EM and GW selection of multi-messenger events. We then outline the analysis framework used to infer the BH and NS mass distributions and the nuclear EOS from the simulated samples.
We present a new methodology which is also able to provide constraints on the magnitude, length scale, and location in energy density of structure in the EOS. 
We conclude by discussing improvements in the analysis chain that would be required in order to obtain reliable constraints from real multi-messenger observations of NSBH mergers.

\begin{figure*}
    \centering
    \includegraphics[width=\textwidth]{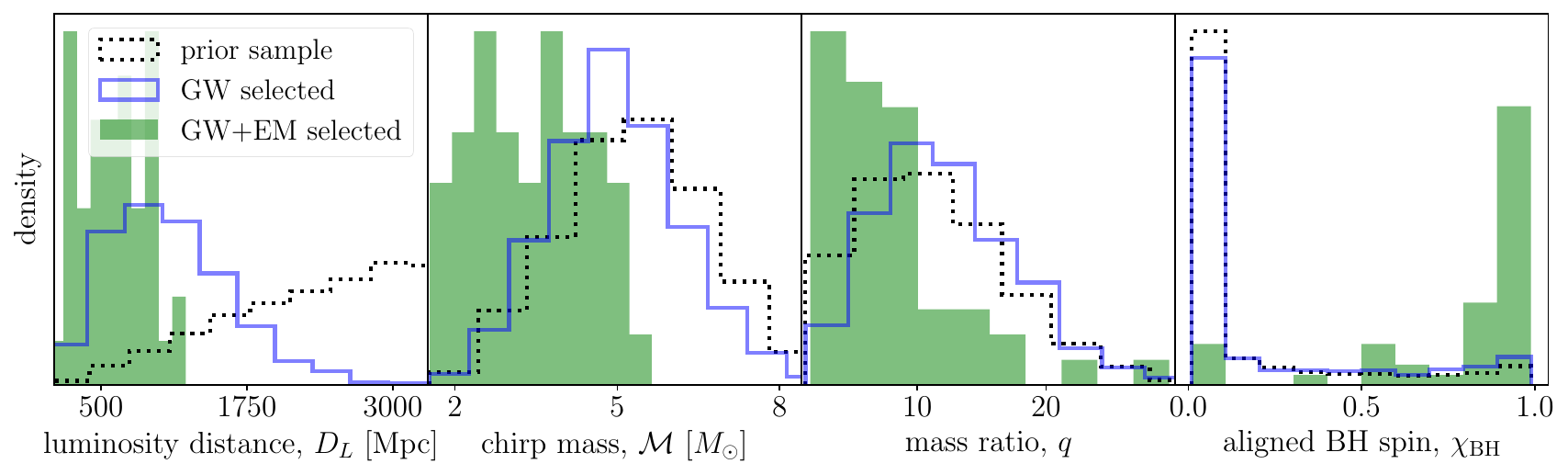}
    \caption{Distributions of a subset of parameters for our simulated population. In particular, luminosity distance $D_{L}$, chirp mass $\mchirp$, mass ratio $q=\mbh/\mns$, and the black hole spin $\chi_{\rm BH}$ of the binary. The black curves show the full simulated population, the blue curves show the GW selected population, and the green histograms show the multi-messenger (EM + GW) selected population.}
    \label{fig:population}
\end{figure*}

\section{Simulations}
We start by defining a population model of NSBH systems motivated both by stellar population synthesis simulations~\citep{Broekgaarden2021} and current astrophysical constraints~\citep{gwtc3pop}. 
We assume a constant (non-evolving) NSBH merger rate of $\Gamma = 32$ $\yrgpc$, which is median value obtained from the the third LVK GW catalog using a Binned Gaussian Process model~\citep{gwtc3pop}. The NSBH mergers are distributed uniformly in comoving volume and isotropically over the sky. We assume truncated normal mass distributions: $P(\mbh / M_{\odot}) = \mathcal{N}(\mu=15, \sigma=10, a=2.42, b=55)$; and $P(\mns / M_{\odot}) = \mathcal{N}(\mu=1.5, \sigma=0.3, a=1.1, b=2.1)$. We adopt the DD2 EOS~\citep{Hempel2010} and use the maximum implied NS mass as the lower limit for the BH mass distribution. However, we restrict the NS mass distribution to be truncated at a lower mass as these quantities likely differ due to NSBH formation processes~\citeg{Chattopadhyay2021}.  
For the distribution of BH and NS spins, which we assume to be aligned, we adopt $P(\chi_{\rm BH}) = \rm{Beta}(\alpha=0.1, \beta=0.5)$ and $P(a_{\rm NS}) = \rm{Uniform}(0, 0.05)$, respectively. A beta distribution is consistent \citep{Chattopadhyay2021} with the predictions from population synthesis simulations when post-processed using the prescription from Ref.~\citep{Qin2018}. We have independently confirmed this finding using the same approach, verifying that post-processing simulations from Ref.~\citep{Broekgaarden2021} produces an asymmetrical BH spin distribution consistent with our beta distribution prior. We set the BH tidal deformability to $\Lambda_{\rm BH} = 0$; $\Lambda_{\rm NS}$ is set by the component masses and the DD2 EOS. 

We simulate observations for a $5$-detector GW network at O5-sensitivities, expected to operate from 2027~\citep{abbott20_LRR}. 
This network consists of two LIGO A+~\citep{LIGO}, Advanced Virgo~\citep{Virgo}, KAGRA~\citep{Kagra}, and LIGO India~\citep{ligo_india, abbott20_LRR}, for which sensitivity curves are available~\footnote{Sensitivity curves from  \url{dcc.ligo.org/LIGO-T2000012/public}. For KAGRA and Advanced Virgo we use the optimistic, $\unit[128]{Mpc}$ and ``high'' range sensitivity curves, respectively.}. We assume an observing time of $\tobs = \unit[5]{yrs}$ with a duty cycle of $\fobs = 0.5$.

We begin by drawing the total number of mergers from a Poisson distribution with mean $\lambda = 4\pi / 3 \, \fobs  \, \tobs \, \dlmax^{3} \, \rate$, where $\dlmax{}=\unit[3,500]{Mpc}$ is the highest luminosity distance at which the most massive NSBH merger could be detected in O5 for our population, consistent with horizon distance measurements of NSBH mergers~\citep{Gupta2023}. The realization we analyze has $14,608$ mergers within a sphere of radius $\dlmax{}$ over the 5-year observing period. 

For each merger, we generate mock data by creating a GW signal with the \waveform{} waveform~\citep{Matas2020}, and injecting the signal into the $5$-detector GW network described above, creating at most a $\unit[160]{s}$ signal for our simulated events for a frequency range of $20$--$2048\, \unit{Hz}$. 
For each event, we calculate the matched-filter network GW signal-to-noise ratio (SNR), $\rho_{\rm MF}$, considering a signal detectable in GWs if $\rho_{\rm MF} \geq 12$. 
The GW selection threshold is passed by $1,392$ of the $14,608$ simulated mergers.
We also calculate the mass disrupted during the merger following Refs.~\citep{Foucart_2018, Zappa2019}, with the relations for calculating the disk mass in NSBH mergers calibrated to numerical simulations performed with the DD2 EOS, 
assuming $30\%$ of the disk is ejected. 
Models of kilonova emission from NSBH mergers are not yet well understood as, e.g., there is significant uncertainty about the electron fraction of the ejecta and the quantity of dynamical and disk-wind ejecta. In view of these uncertainties, we assume any merger with disk ejecta of $\gtrsim 0.01 M_{\odot}$ will produce a detectable kilonova within $\unit[500]{Mpc}$ and use this reference mass and distance to build an EM selection function. The choice of reference mass and distance is consistent with projections for the detectability of kilonovae across a range of numerical simulations of different kilonovae properties in optical surveys such as the Vera C. Rubin Observatory's Legacy Survey of Space and Time (LSST)~\citep{Ivezic2019, Andreoni2022}. 
Of our GW-selected mergers, $95$ events have disk ejecta masses $M_{\rm ej} \geq 0.01 M_{\odot}$; then applying the secondary distance selection criterion leaves a final sample of $47$ events that pass GW and EM selection (from the 14,608 initially simulated). Of these, there are $37$ multi-messenger events with disk masses $\gtrsim 0.1 M_{\odot}$, comparable to the disk mass inferred for GW170817~\citep{Radice2018} and likely sufficient to launch a relativistic jet which could be observable as a gamma-ray burst with a broadband afterglow~\citep{Hinderer2019, Raaijmakers2021_mm, Barbieri2019, Sarin2022_linking}.

We include EM selection as EM emission provides definitive evidence of disruption, allowing us to be confident that a  system is an NSBH instead of a BBH, ruling out EM emission from a BNS, and therefore yielding a tighter constraint on the tidal deformability. (EM plays no other role in our analysis beyond ensuring the purity of the NSBH sample.)
For our full population, we expect $7\%$ of mergers to produce EM emission in the form of a relativistic jet and/or a kilonova, consistent with current constraints~\citep{Sarin2022_linking, Biscoveanu2023}. 
While our EM selection treatment does not account for the diversity of brightness and color of different mergers~\citep{Barbieri2019, Setzer2019}, viewing-angle dependence~\citep{Zhu2020} or the effect of survey cadences~\citeg{Setzer2019, Almualla2021}, and is only calibrated to simulations performed with the DD2 EOS\footnote{See~\citet{Henkel2023} for discussions on the agreement between different ejecta models).}, we expect our threshold on a reference distance and ejecta mass to capture the critical features of the selection function. 

We present the impact of GW and EM selection on our population in Fig.~\ref{fig:population}. 
The predominant effect of the GW selection is to favor nearby mergers and face-on events, both of which produce a stronger GW signal. By contrast, the most significant impact of EM selection is on luminosity distance, mass ratio, chirp mass and BH spin, with lower chirp masses and higher spins leading to more favorable conditions for disrupting the NS in order to produce a detectable EM transient; such events must also be sufficiently nearby to be detectable.

\section{Analysis methods}
We use a Bayesian hierarchical model (BHM) to constrain the NS and BH mass functions and the NS EOS, analyzing these jointly to avoid biases that can arise from estimating each individually~\citep{Golomb2022}. The posterior distribution of these population parameters, $\Omega$, is obtained by also inferring the object-level parameters of the $N$ detected mergers, $\theta_{1:N}$, and then marginalizing over these (along with any population parameters that are not of direct interest, such as the overall rate normalization). Assuming an uninformative prior on the normalization, the marginal posterior on the other population parameters can be written as \cite{Mandel2019, Vitale2022}
\begin{equation}
\label{equation:post}
P(\Omega | d_{1:N}, I)
\! \propto \!
P(\Omega | I) \, 
\frac{
\prod_{i = 1}^N
\int {\rm d} \theta_i \,
P(\theta_i | \Omega) \, 
P(d_i | \theta_i)
}
{
[P(S | \Omega)]^{N}
}
,
\end{equation}
where $P(\Omega | I)$ is the population-level prior, given our our prior background information $I$, 
$P(S | \Omega)$ is the (EM and GW) selection probability averaged over the population, and $d_{1:N}$ is the GW data for the $N$ detected mergers.
We approximate the selection and marginalization integrals using a two-step approach: we first perform individual object-level inference using reference values of the global parameters, $\Omega_0$; and we then use importance resampling to combine these results to constrain $\Omega$.

For each event we take the object-level parameters, $\theta$ to be the standard aligned spin parameter set~\citep{romeroshaw20}.  For the $i$'th selected event (with $i \in \{1, 2, \ldots, N\}$) we sample the posterior distribution $P(\theta_i | d_i, \Omega_0 )$ using the ensemble sampler {\sc{emcee}}~\citep{Foreman-Mackey2013} as implemented in \bilby{}~\citep{bilby,romeroshaw20}. 

We explored \program{dynesty}~\citep{Speagle2020}, \program{bilby\_mcmc}~\citep{Ashton2021}, and \program{nessai}~\citep{Williams:2021qyt} samplers through \bilby{} for parameter estimation on individual events across our sample. Although these samplers have been tested for various GW data analysis tasks, our stringent setting of a five-detector network operating at O5 sensitivities in combination with a relatively expensive, effective one-body waveform, \waveform{} proved challenging. 
We encountered consistent problems with parameter recovery across the population as well as issues with convergence, coupled with high computational costs for the analysis. Furthermore, techniques such as relative binning~\citep{Zackay2018}, which have been demonstrated to dramatically reduce wall-clock time of parameter estimation on single events, also failed to provide converged results across our sample, which we tracked down to inadequacy of the likelihood approximations involved. To bypass these issues, we used \program{emcee}, starting the ensemble walkers within a narrow volume around each event's true parameters. This setting ensured convergence across the population and reduced the wall-clock time of the analysis. We ensured that our results with \program{emcee} are robust by performing multiple runs for each event and that we obtained consistent posteriors for events without convergence issues obtained by our analysis with different samplers. While this approach is sufficient for the purpose of this simulation study, our experience highlights the upgrades to the analysis framework, particularly related to sampling, that will be required for upcoming population studies with next-generation GW data.

The standard reference model used in GW inference assumes an EOS-agnostic uniform prior on the two tidal component deformability, i.e., $P(\Lambda | \Omega_0) = \rm{Uniform}(0, 5000)$, ignoring information provided by the mass ratio of the binary or that all EOS forms predict that $\Lambda(m)$ is a smoothly decreasing function of $m$. Our chosen waveform model, \waveform{}, can only be evaluated for $\Lambda_1 = 0$~\citep{Matas2020}, i.e., implicitly assuming that the primary component is known to be a BH, something we assume can be ensured through coincident EM observations. 
This assumption could be relaxed by choosing a BNS waveform and allowing the data to dictate the measurement, but such waveforms are not calibrated to NSBH simulations and are not designed to work for the range of mass ratios of such NSBH systems~\citep{Dietrich2019}, which could bias results. 
We therefore use the \waveform{} waveform as it is built on the effective-one-body formalism to model the two-body problem in general relativity and calibrated to numerical NSBH simulations~\citep{Matas2020}. The resulting posteriors in NS mass and tidal deformability for all 47 detected multi-messenger events are shown in the Supplemental Material. We do not fix any parameter from the standard GW aligned spin parameter set apart from $\Lambda_1$ to the true input value. Each object-level inference analysis takes up to $\unit[3]{days}$ on an Intel Xeon 6140 CPU. 

The individual single-event posteriors for all events can now be combined. We first construct a continuous representation of our single event likelihoods using Gaussian mixture models (GMMs) with three components. This requires transforming the original posterior samples into a better-suited domain~\citep{Golomb2022}. 
The result is an approximate likelihood  $\tilde{P}(d_i | \theta, \Omega)$ valid for reasonable parameter values for each of the $N$ mergers.
The selection probability is estimated by simulating $K \gg N$ mergers under the reference model $\Omega_0$ and recording the parameters $\theta_{1:J}$ for the $J \leq K$ mergers which satisfy both the EM and GW selection. 
The marginalized posterior in Eq.~\ref{equation:post} can then be approximated as 
\begin{equation}
\label{equation:mcint}
P(\Omega | d_{1:N}, I)
\! \propto \!
\prob(\Omega | I)
\, \frac{ \prod_{i=1}^N
1/S \, 
\sum_{s = 1}^S 
\tilde{P}(d_i | \theta_s)
}
{
[1 / K \, \sum_{j = 1}^J 
P(\theta_j | \Omega) / P(\theta_j | \Omega_0)]^N
}
,
\end{equation}
where $\theta_{1:S}$ are $S$  draws from the prior $\prob(\theta | \Omega)$.
We find $S = 20,000$ samples sufficient for convergence. 

\begin{figure}
      \centering
\includegraphics[width=\columnwidth]{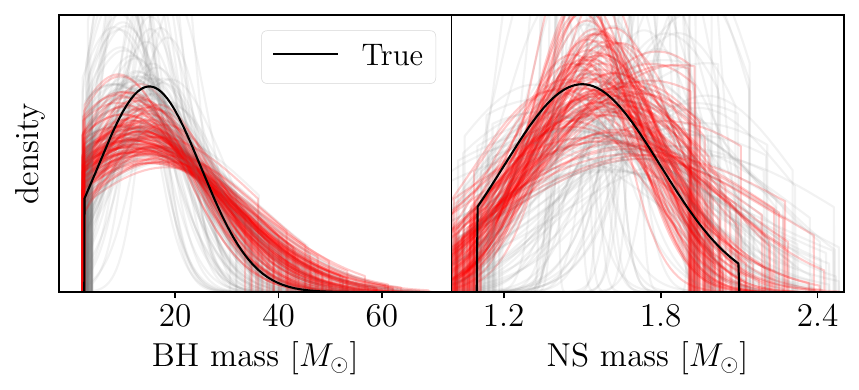}
    \caption{Constraints on the BH mass distribution (left) and the the NS mass distribution (right). The grey and red curves are draws from the prior and posterior, respectively. The black curve is the true input of our simulation.}
    \label{fig:hyperppdmass}
\end{figure}

\begin{figure*}
    \centering
    \includegraphics[width=\textwidth]{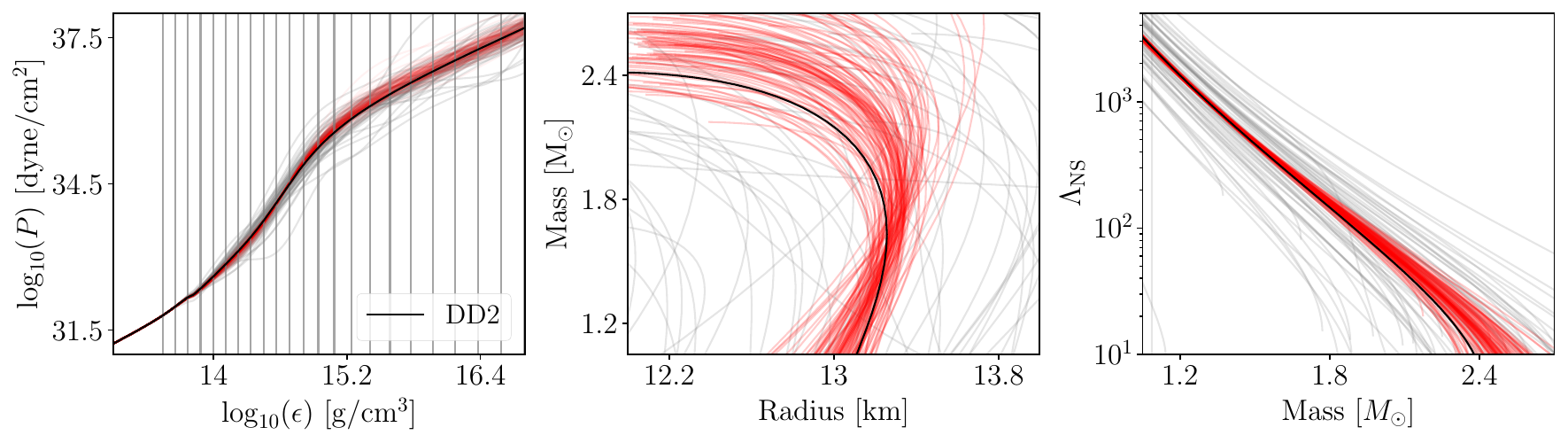}
    \caption{Constraints on (from left to right) the nuclear EOS, the NS mass-radius relationship and and the mass-$\Lambda$ relationship. The grey and red curves are draws from the prior and posterior, respectively. The black curve is the true input of our simulation. The gray vertical lines indicate the locations of our $24$ nodes.}
    \label{fig:hyperppdeos}
\end{figure*}

Our population model parameters, $\Omega$, describe the mass distributions of BHs and NSs and the nuclear EOS. For the former, we use truncated normal distributions with the same parameters used in our simulations. The prior distributions for these parameters are listed in Table~\ref{table:priors}. 

The prior for the EOS must be chosen more carefully, being defined on the space of (monotonically increasing) functions $p(\epsilon)$ which encode the dependence of pressure, $p$, on energy density, $\epsilon$. For numerical calculations we work with the logarithms of these two quantities, expressing $\epsilon$ in units of $\unit{MeV~fm^{-3}}$ and $p$ in units of $\unit{dyne~cm^{-2}}$. Simply adopting common flexible parameterizations such as a piece-wise polytrope~\citep{Read2009} or spectral-decomposition~\citep{Lindblom2010} can, however, introduce undesirable implicit correlations~\citep{Raaijmakers2018_pitfalls, Legred2022}. We hence build a more flexible Gaussian process (GP) prior for the (log) EOS~\citep{Landry2019, Essick2020, Legred2022}. This is defined by $n$ nodes, $\log_{10}(\epsilon_{1:n})$ and $\log_{10}(p_{1:n})$, indexed by $k \in \{1, 2, \ldots, N\}$. This approach is strictly valid only in the $n \rightarrow \infty$ limit, but for the nuclear EOS we find that $n = 24$ is sufficient for practical purposes. The positions of the nodes in log energy density, $\log_{10}(\epsilon_{1:n})$, are shown in Fig.~\ref{fig:hyperppdeos}. For the prior mean of the (log) pressure of the $k$'th node we set $\log_{10}(\bar{p}_k) = \log_{10}[p_{\rm DD2}(\epsilon_k)]$ as given by the DD2 EOS. Below an energy density of $\unit[19.7]{MeV~fm^{-3}}$, we force the GP to match DD2, i.e., conditioning the GP prior to match the better-known low density physics of the NS crust~\citep{Hempel2010} (as done in other EOS analyses~\citeg{Essick2020}). 
To encode the correlations between the pressure at different nodes we adopt a squared exponential GP kernel of the form $C_{k,k^\prime}=k_1 \exp\{-[\log_{10}(\epsilon_k) - \log_{10}(\epsilon_{k^\prime})]^2/(2\,k_2^2)\}$, where the amplitude, $k_1$, and scale, $k_2$, become global nuisance parameters which are constrained by the node-to-node covariance of the simulated EOS.

We sample an EOS from this prior distribution by using a three step process. We first draw $\log_{10}(\mathbf{p}) = \log_{10}(p_{1:n})$ from an $n$-dimensional multivariate normal distribution\footnote{A simple way of generating a random draw $\mathbf{x}$ from a multivariate normal distribution of mean $\boldsymbol{\mu}$ and covariance $\mathbf{C}$ is to i) generate a random vector $\mathbf{n}$ with elements drawn from a unit normal and then ii) set $\mathbf{x} = \boldsymbol{\mu} + \mathbf{L} \, \mathbf{n}$, where $\mathbf{L}$ is the Cholesky decomposition of $\mathbf{C}$ (i.e., a lower triangular matrix such that $\mathbf{L} \, \mathbf{L}^{\rm T} = \mathbf{C}$).} with mean $\log_{10}(\bar{\mathbf{p}}) = \log_{10}(\bar{p}_{1:n})$ and covariance matrix $\mathbf{C} = C_{1:n,1:n}$. 
We then form a full EOS by using GP interpolation to transform the $\log_{10}[p(\epsilon)]$ at $n=24$ nodes onto a denser array of $2000$ nodes.
This can produce models that are acausal or thermodynamically unstable~\citep{Lattimer2001}; these are removed using rejection sampling. For the implied $p(\epsilon)$, we solve the tidal and Tolman-Oppenheimer-Volkoff (TOV) equations to obtain $\Lambda(m)$, which is used in the evaluation of the Monte Carlo integral in the likelihood in Eq.~\ref{equation:mcint}.

\begin{table}[h!]
\resizebox{\columnwidth}{!}{
\begin{tabular}{||l|c|c||} 
 \hline
 Definition & Parameter & Prior \\ [0.5ex] 
 \hline\hline
BH mass function mean & $\mu_{\rm BH} / \unit{M_{\odot}}$ &  $\uniform(8,20)$\\ 
BH mass function width & $\sigma_{\rm BH} / \unit{M_{\odot}}$ & $\uniform(4,20)$ \\
BH mass function minimum & $a_{\rm BH} / \unit{M_{\odot}}$ & $\uniform(2,4)$ \\
BH mass function maximum & $b_{\rm BH} / \unit{M_{\odot}}$ & $\uniform(30,70)$ \\
NS mass function mean & $\mu_{\rm NS} / \unit{M_{\odot}}$ &  $\uniform(1.3,2.0)$\\ 
NS mass function width & $\sigma_{\rm NS} / \unit{M_{\odot}}$ & $\uniform(0.01,0.6)$ \\
NS mass function minimum & $a_{\rm NS} / \unit{M_{\odot}}$ & $\uniform(0.9,1.4)$ \\
NS mass function maximum & $b_{\rm NS} / \unit{M_{\odot}}$ & $\uniform(1.9,2.5)$ \\
EOS GP amplitude & $\log(k_1)$ &  $\uniform(0.1, 0.8)$\\ 
EOS GP scale & $k_2$ &  $\uniform(0.2, 1.0)$\\ [1ex]
 \hline
\end{tabular}}
\caption{Population parameters and their prior distributions.}
\label{table:priors}
\end{table}

With this EOS parameterization, we have a total of 34 population parameters: the parameters of the BH and NS mass distributions; the GP kernel hyper-parameters; and the pressure at the $n = 24$ EOS nodes. In Fig.~\ref{fig:hyperppdmass} and Fig.~\ref{fig:hyperppdeos}, we show random draws from our full prior for different NS and BH properties. Our priors are summarized in Table~\ref{table:priors}.
We obtain the posterior on these hyperparameters using the likelihood in Eq.~\ref{equation:mcint}, sampling from $P(\Omega | d_{1:N}, I)$ using the nested sampler {\sc{pymultinest}}~\citep{Feroz2009, Buchner2016} implemented in \bilby{}~\citep{bilby, romeroshaw20}. 
We evaluate the likelihood on a GPU using {\sc{cupy}} to reduce the computational cost. 
The analysis steps from transforming the individual event posteriors, building a Gaussian mixture model density estimate, to producing posteriors on the hyperparameters takes $\mathcal{O}(\unit[1]{day})$ on a NVIDIA P100 GPU, limited primarily by the need to solve the TOV equations at every iteration of the likelihood.

\section{Results}
Following the methodology outlined above, we present the results from our EMGW-selected population of NSBH mergers. Our simulation recovers the input values for all parameters, indicating no bias in our analysis and that we have correctly accounted for selection effects. 

In Fig.~\ref{fig:hyperppdmass} and~\ref{fig:hyperppdeos}, we show the prior and posterior predictive distributions for the BH mass distribution $P(\mbh)$, the NS mass distribution $P(\mns)$, the NS EOS (i.e., $p(\epsilon)$), and the mass-radius and the mass-$\Lambda$ curves. 
We again see that the true input of the simulation is recovered, indicating no bias in our analysis and correct accounting of selection effects in this projected representation of parameters.
We measure the BH and NS mass distribution means with a precision of $46\%$ and $27\%$, respectively, at the $95\%$ credible interval.
However, the high mass cut-off in the NS mass distribution and low-mass cut-off in the BH mass distribution are not constrained well, with significant overlap suggesting that a sample of this size will not be able to verify the existence of a mass gap between NSs and BHs, consistent with previous results~\citep{Alsing2018, Ye2022}.

To quantify the constraining power on the nuclear EOS, we can consider the constraints on the tidal deformability and radius of a $\unit[1.4]{M_{\odot}}$ NS as $\lambdaconstraint$ (${765.93}_{-71.70}^{+141.43}$) and $\radconstraint~\rm{km}$ (${13.41}_{-0.34}^{+0.39}~\rm{km}$) for a $68\%$ ($95\%$) credible interval, i.e., a precision of $10\%$ and $1.6\%$ for a $68\%$ credible interval respectively. Similarly, we can also constrain the maximum NS mass to be ${2.50}_{-0.14}^{+0.19} M_{\odot}$ ($68\%$ credible interval), i.e., a relative precision of $13\%$. The precision of each measurement is comparable to other state-of-the-art methods to constrain the behaviour of nuclear matter~\citeg{Riley2019, Raaijmakers2019, Essick2020}, demonstrating the importance of constraints provided by observations of NSBHs. 

Further, a significant benefit of our new GP-based EOS inference methodology is that it directly constrains the pressure at specific energy densities (the GP nodes) and the size and length scale parameters of correlations in $p(\epsilon$). Our simulations imply that the length scale of correlations (i.e., the smoothness) in $p(\epsilon)$ can be constrained to be $\geq~\unit[3.2]{MeV~fm^{-3}}$ with $90\%$ confidence, an important consideration for determining the size and location of putative phase transitions.  

Given the local rate of NSBH mergers is highly uncertain~\citep{gwtc3pop}, we also perform our analysis assuming a pessimistic rate of $8~ \yrgpc$,  the lowest rate estimate from the Binned Gaussian Process model~\citep{gwtc3pop}. We redo our full analysis using only a random sample of $11$ events from our multi-messenger selected population. This reduced sample yields constraints on the tidal deformability and radius of a $\unit[1.4]{M_{\odot}}$ NS of ${845.72}_{-127.24}^{+218.95}$ and ${13.61}_{-0.38}^{+0.55}~\rm{km}$ ($68\%$ credible intervals), respectively. This indicates that the precision from a quarter of the population on the radius and tidal deformability of a $\unit[1.4]{M_{\odot}}$ will be approximately four times worse. However, the EOS constraints are dominated by the high SNR events~\citeg{HernandezVivanco2019}; so while the scaling of constraints presented above is in a typical scenario, there is always the possibility of a fortuitous high SNR event which allows reaching a specific constraint more quickly.
\section{Conclusions}
We have presented the constraints on the BH and NS mass distributions and the nuclear EOS that could be provided by a sample of multi-messenger NSBH events from $\unit[5]{years}$ of the A+ era GW observatories operating in tandem with large-scale optical surveys like LSST. Our EOS constraints come only from the GW data, with EM selection only serving to ensure a pure NSBH sample. Folding in the EOS dependence into the modeling of the EM counterpart could further improve constraints from such mergers~\citep{Raaijmakers2021_mm}. Our EOS inference methodology also offers the ability to directly constrain structure in the EOS, an important consideration for probing the existence of phase transitions. 

The precision of EOS constraints provided by such a sample of NSBH mergers are comparable to projected constraints from BNS mergers and better than constraints provided by NICER~\citep{Raaijmakers2019}. In particular, for $47$ multi-messenger NSBH events, we can obtain $1.6\%$ precision measurement on the radius of $\unit[1.4]{M_{\odot}}$ NS cf. $2\%$ constraint for $\sim 50$ BNS mergers for a similar equation of state and 3-detector GW network at design sensitivity~\citep{Finstad2022}. Currently, the local rate of both BNS and NSBH mergers are highly uncertain~\citep{gwtc3pop}. 
However, the number of NSBH candidates currently outnumber BNS candidates, and this could conceivably continue given the former are detectable out to a larger volume. This study demonstrates the strong complementarity of NSBH mergers as a probe of the behavior of nuclear matter, especially given that it is unclear as yet which merger type will dominate future EMGW samples.  

A number of improvements will be required to realize the promise of NSBH mergers. For example, the analysis of real observations will require a more sophisticated treatment of EM selection that incorporates viewing angle dependencies, the intrinsic diversity of EM counterpart signals~\citep{Setzer2023}, and real survey observing strategies~\citep{Andreoni2022}, alongside improvements to physical models of EM counterparts to ensure that kilonovae from NSBH can be robustly identified. In particular, several improvements are required for kilonovae modeling, such as improved prescriptions and a better understanding of nuclear heating~\citep{rosswog24a}, ejecta opacities~\citep{tanaka14a} and the precise ejecta properties of NSBH mergers and how they link to the progenitor system~\citeg{Foucart_2018}.
If GW observations alone could ensure a pure NSBH sample (i.e., by ruling out contamination from BNS or BBH mergers)~\citeg{Brown2022}, this would remove the need for EM selection, and the systematics associated with kilonova models, dramatically increasing the number of observations available to constrain the EOS. From the perspective of gravitational-wave data, the role of a number of other systematics must also be better understood. In particular, this includes the systematic uncertainty from the choice of population model for black hole and neutron star masses, and spins~\citep{Wysocki2020, Golomb2022}, bias due to physics potentially not included in event-level analyses such as higher-order modes~\citep{veske21} or eccentricity~\citep{divyajyoti24}.
Further, as constraints on the EOS are dominated by events with high SNR, a better understanding of waveform systematics in that regime will be essential~\citeg{Purrer2020, Gamba2021, Huang2021}.
Finally, it may be promising to investigate building more physical relationships (such as density-dependent correlations seen in numerical EOS models) into EOS priors, while retaining the advantage of the flexibility offered by GP modeling.

\section*{Data and code availability}

This work used \bilby{}~\citep{bilby}, available at \url{https://git.ligo.org/lscsoft/bilby} and \program{Redback}~\citep{sarin23_redback}, available at \url{https://github.com/nikhil-sarin/redback}. Specific analysis scripts are available at \url{https://github.com/nikhil-sarin/nsbh-eos-analysis}.
\section*{Author contributions}

{\bf NS}: conceptualization, methodology, software, investigation, validation, writing (original draft preparation). {\bf HVP}: conceptualization, methodology, validation, writing (review \& editing) {\bf DJM}: conceptualization, methodology, writing (review \& editing). {\bf JA}: methodology, validation, writing (review \& editing). {\bf SMN}: conceptualization, writing (review \& editing). {\bf SMF}: conceptualization. 

\section*{acknowledgements}
We thank Alex Brown,  Eric Thrane, Greg Ashton, and Shanika Galaudage for helpful discussions. 
We are grateful to the \texttt{SEOBNR} waveform modelers for making their waveforms public, without which this work would not have been possible.
NS is supported by a Nordita Fellowship. Nordita is supported in part by NordForsk. This project has received funding from the European
Research Council (ERC) under the European Union’s Horizon 2020 research and innovation programmes (grant agreement no. 101018897 CosmicExplorer) and by the research project grant `Fundamental physics from populations of compact object mergers' funded by VR under Dnr 2021-04195. This work has benefitted from the research environment grant `Gravitational Radiation and Electromagnetic Astrophysical Transients (GREAT)' funded by the Swedish Research Council (VR) under Dnr 2016-06012 and the research project grant `Gravity Meets Light' funded by the Knut and Alice Wallenberg Foundation Dnr KAW 2019.0112. The work of HVP was additionally supported by the Göran Gustafsson Foundation for Research in Natural Sciences and Medicine. SMN is grateful for financial support from the Nederlandse Organisatie voor Wetenschappelijk Onderzoek (NWO) through the VIDI and Projectruimte grants. NS, HVP and DJM acknowledge the hospitality of the Aspen Center for Physics, which is supported by National Science Foundation grant PHY-1607611. The participation of HVP and DJM at the Aspen Center for Physics was supported by the Simons Foundation. This work used computing facilities provided by the OzSTAR national facility at Swinburne University of Technology. The OzSTAR program receives funding in part from the Astronomy National Collaborative Research Infrastructure Strategy (NCRIS) allocation provided by the Australian Government. This material is based upon work supported by NSF's LIGO Laboratory which is a major facility fully funded by the National Science Foundation.


\appendix
\section{Supplemental Material}
\begin{figure*}
\centering
    \includegraphics[width=\textwidth,height=24cm]{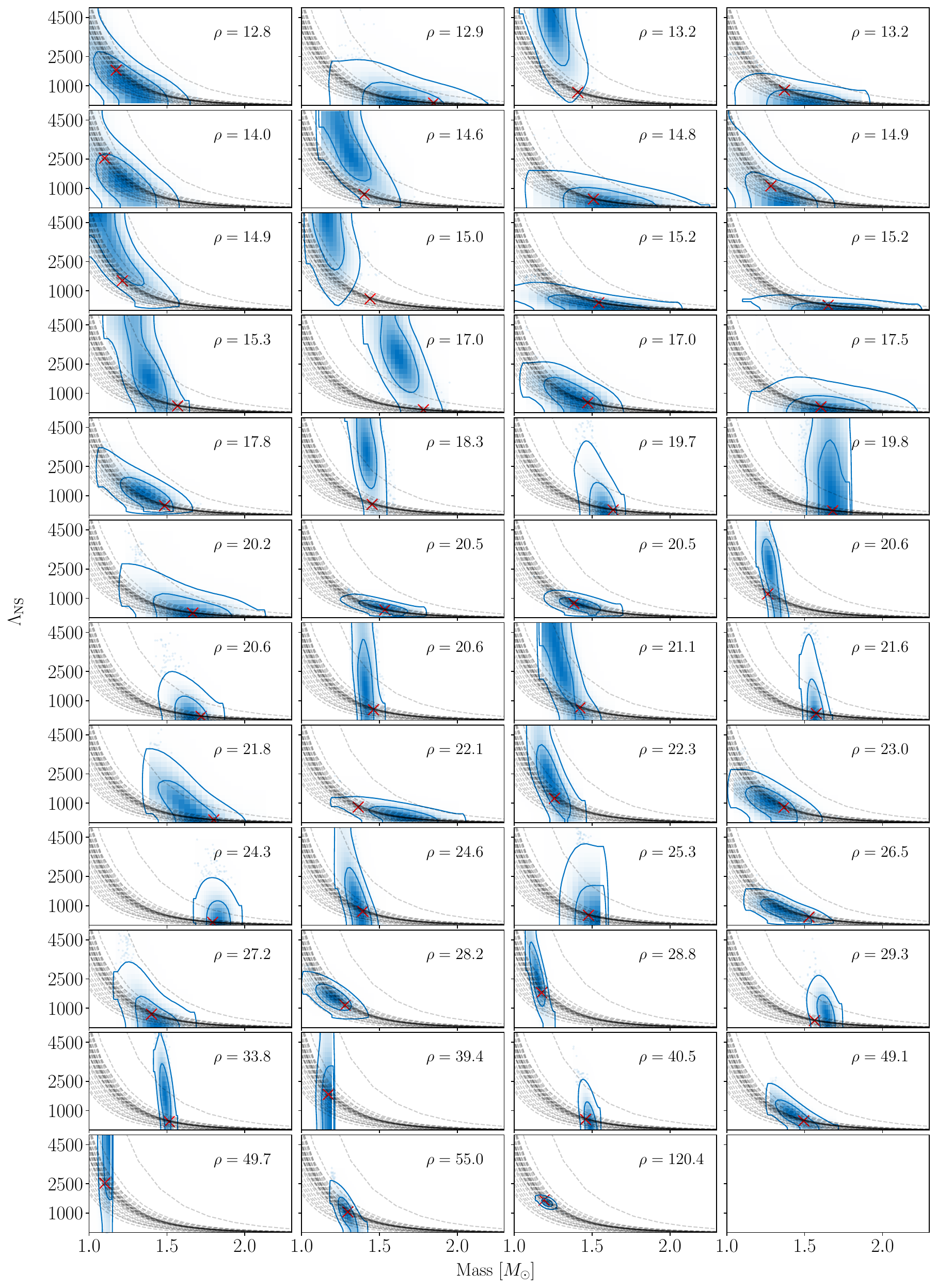}
    \caption{Constraints on the mass and tidal deformability of neutron stars in our all events with the blue shaded region indicating the $50$ and $90\%$ credible regions and the red cross showing the input value. The annotation indicates the network matched-filter SNR of our events, while the gray curves indicate random draws from the GP prior.}
    \label{fig:singleposteriors}
\end{figure*}

\textbf{\emph{Individual event posteriors.}} Here we present the posteriors from our individual event analysis and from the full population analysis. In Fig.~\ref{fig:singleposteriors} we show the constraints on the mass and radius from the individual events. The $3\sigma$ credible regions for all events include the true/input value, giving confidence in an unbiased recovery. Fig.~\ref{fig:singleposteriors} also demonstrates that the dominant constraints on the $\Lambda(m)$ curve are provided by the loudest GW events, consistent with previous work~\citep{HernandezVivanco2019, Golomb2022}. This highlights that the constraints on the nuclear EOS will be highly dependent on the ability to detect a handful of exceptionally loud events, as opposed to many weak events, stressing the need for understanding waveform systematics for high SNR systems~\citeg{Purrer2020}. 

\bibliography{paper}

\end{document}